\documentclass[12pt]{article}%
\usepackage{amsmath}
\usepackage{graphicx}
\usepackage{amsfonts}
\usepackage{amssymb}%
\providecommand{\U}[1]{\protect\rule{.1in}{.1in}}
\begin{document}

\title{Delta-Interference of Two Friedel Resonances}
\author{Gerd Bergmann\\Department of Physics\\University of Southern California\\Los Angeles, California 90089-0484\\e-mail: bergmann@usc.edu}
\date{\today }
\maketitle

\begin{abstract}
When a single resonator is coupled to a continuous spectrum one obtains a
resonance of finite half-width. Such a resonance is known in many fields of
physics. The Friedel resonance is an example where a d-impurity is dissolved
in a simple metal. If two resonators are coupled to the continuous spectrum
the resonances interfere. For identical coupling and frequencies one obtains
two effective resonances. The effective coupling of one of them to the
continuum can be tuned to zero yielding a delta-like resonance.

\end{abstract}

Sharp resonances provide important markers in physics. For example in solid
state physics the sharp Kondo resonance has been used in a number of beautiful
experiments to observe the propagation of electrons in real space \cite{E18},
to see the Fermi surface of the host \cite{W47} and to measure magnetic energy
shifts \cite{H50}. In this paper I propose to build a sharp resonance out of
two broad resonances. The ability to fine-tune the strength and width of the
resonance will be very useful in similar experiments.

A well known example of a broad resonance in solid state physics is the
Friedel resonance. It forms when a d-impurity is dissolved in a simple
(s,p)-metal \cite{F28}. The d-level hybridizes with the conduction electrons.
An electron injected into the d-level experiences a finite lifetime which
results in a finite finite half-width of the d-state. The latter is given by
the golden rule
\[
\Delta=\pi\overline{\left\vert V_{sd}\right\vert ^{2}}\rho_{0}%
\]
where $\overline{\left\vert V_{sd}\right\vert ^{2}}$ is an average over the
s-d-hopping matrix element and $\rho_{0}$ is the density of states. Similar
resonances have been investigated by Feshbach \cite{F51} and Fano \cite{F31}.
These resonances play an important role in many areas of physics, such as
nuclear and atomic physics as well in solid state.

When a finite concentration of d-impurities is dissolved in the host one
generally assumes that their resonances are independent and additive. The main
justification is that the matrix elements $V_{sd}$ have different phases for
different positions. Take as a very simple example a one-dimensional wire of
length $L$ with periodic boundary conditions. Then the electron wave functions
can be expressed as $\varphi_{k}\left(  x\right)  =\sqrt{1/L}\cos\left(
kx\right)  $ or $\sqrt{2/L}\sin\left(  kx\right)  $. If we put one impurity at
$x=0$ then it has the same s-d-matrix element $V$ with all the $\varphi
_{k}\left(  x\right)  =\sqrt{1/L}\cos\left(  kx\right)  $ electrons because
all these states have the same value $\sqrt{2/L}$ at the position of the
impurity. The $\sqrt{2/L}\sin\left(  kx\right)  $-states don't couple to the
d-impurity. If one puts a second d-impurity at $x=L/2$ then the matrix element
alternates between $+V$ and $-V$. Diagonalization of the Hamiltonian shows
that the resonances at $x=0$ and $x=L/2$ just add without distorting each other.

One obtains a completely different result when the matrix elements for the two
resonance levels are identical. How one may achieve this condition will be
discussed below. To broaden the discussion I introduce two d-levels with
different resonance energies $\varepsilon_{d_{1}}$ and $\varepsilon_{d_{2}}$
into an electron system. Both d-impurities shall have identical hopping matrix
elements\ with the conduction electrons. A constant electron density of states
of the host is assumed (which allows a better comparison of resonances at
different energies.) In the numerical calculation 200 electron states with the
energy $\varepsilon_{i}=i/2$ for $0<i\leq200$ are used so that the band
extends from $0$ to $100$. The matrix elements are set equal to $V=1$. In
Fig.1a the resonance energies are well separated at $e_{d_{1}}=25$ and
$\varepsilon_{d_{2}}=75$. One obtains two well separated resonances. The half
widths are $2.5\pi,$i.e. 25\% larger than the theoretical result for a single
Friedel impurity, which is $\Delta=\pi$ $\overline{\left\vert V_{sd}%
\right\vert ^{2}}\rho_{0}=2\pi\ $(in this example the parameters are
$\overline{\left\vert V_{sd}\right\vert ^{2}}=1$ and $\rho_{0}=2$). The full
curves are calculated with this half-width. The \ position of the resonances
are slightly shifted to $e_{d_{1}}^{\prime}=23.8$ and $\varepsilon_{d_{2}%
}^{\prime}=76.2$.
\begin{align*}
&
{\includegraphics[
height=2.6251in,
width=3.1615in
]%
{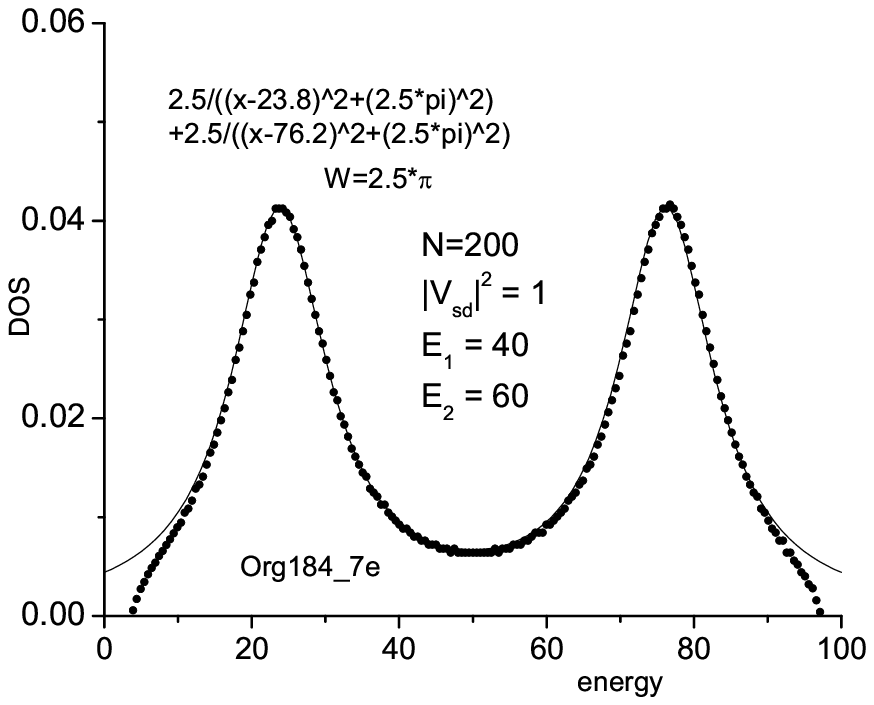}%
}%
\\
&
\begin{tabular}
[c]{l}%
Fig.1: Two d-levels with $e_{d_{1}}=25$ and $\varepsilon_{d_{2}}=75$ with\\
identical coupling to the conduction band
\end{tabular}
\end{align*}
The increased value of $\Delta$ is an indication of an interference between
the two levels.

When the two d-levels are located at $e_{d_{1}}=45$ and $\varepsilon_{d_{2}%
}=55$ one no longer obtains two maxima. The resulting DOS can be described as
two resonances with width of $\Delta_{1}=4\pi$ and $\Delta_{2}=0.7\pi$, both
centered at $\varepsilon_{d}^{\prime}=50$.

If one positions both d-levels at $e_{d_{1}}=\varepsilon_{d_{2}}=50$ then
figure changes dramatically. This is shown in Fig.2.
\begin{align*}
&
{\includegraphics[
height=2.9282in,
width=3.7509in
]%
{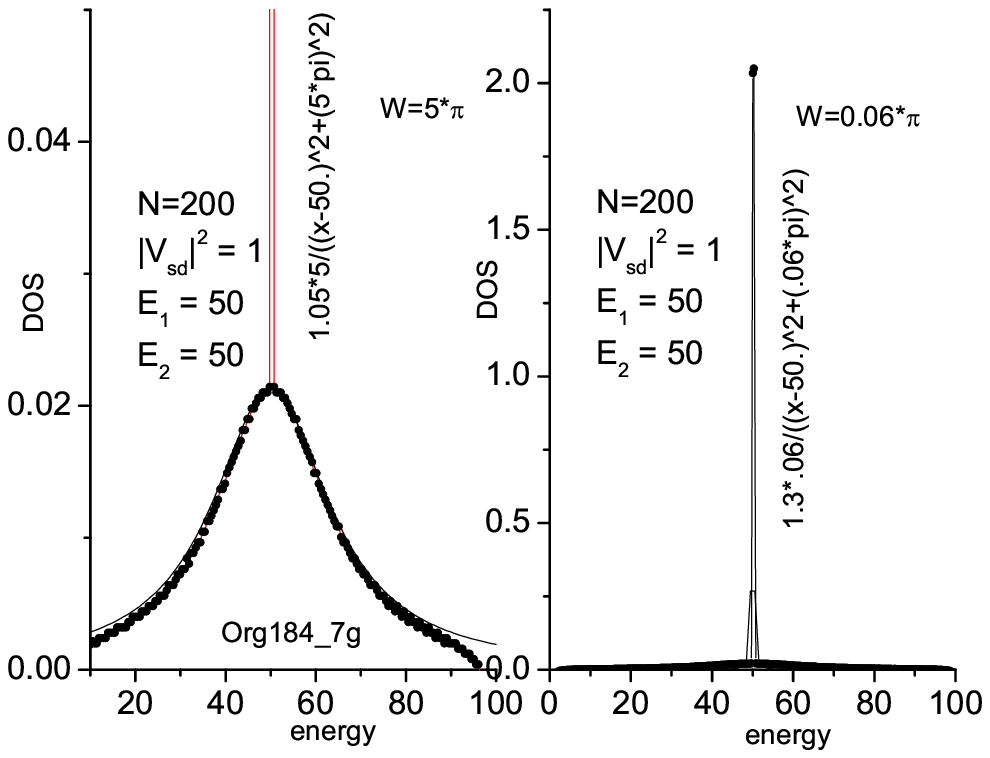}%
}%
\\
&
\begin{tabular}
[c]{l}%
Fig.2: Two identical d-levels with $e_{d_{1}}=$ $\varepsilon_{d_{2}}=50$
with\\
identical coupling to the conduction band. The left figure\\
shows the strong broadening with a needle sharp peek\\
in the center. In the right figure the vertical scale is\\
compressed by two orders of magnitude.
\end{tabular}
\end{align*}
The left figure shows the strong broadening with a needle sharp peek in the
center. The broad resonance has a half width of $\Delta_{1}=5\pi$ and an area
of $1.05$. In the right figure the vertical scale is compressed by two orders
of magnitude. The sharp resonance is very narrow $\Delta_{2}=0.06\pi
\thickapprox\allowbreak0.19$. However, the level separation $\delta
\varepsilon$ of the conduction band is in this calculation only $0.5$ which
prevents a more narrow resonance. As discussed below the resonance width would
approach ideally the value zero.

Since the above result of the simulation is rather surprising and unexpected
it shall be derived by means of Green functions. The GF are defined as
$\mathbf{G=}\left(  E+is-H\right)  ^{-1}$ or
\begin{equation}%
{\textstyle\sum_{\lambda}}
\left(  E+is-H\right)  _{\kappa\lambda}G_{\lambda\mu}=\delta_{\kappa\mu}
\label{GF}%
\end{equation}
with the Hamiltonian%
\begin{gather}
H=%
{\textstyle\sum_{\nu}}
\varepsilon_{\nu}c_{\nu}^{\dag}c_{\nu}+\varepsilon_{d_{1}}c_{d_{1}}^{\dag
}c_{d_{1}}+%
{\textstyle\sum_{\nu}}
\left(  V_{\nu,1}c_{\nu}^{\dag}c_{d_{1}}+V_{\nu,1}^{\ast}c_{d_{1}}^{\dag
}c_{\nu}\right) \label{Ham}\\
+\varepsilon_{d_{2}}c_{d_{2}}^{\dag}c_{d_{2}}+%
{\textstyle\sum_{\nu}}
\left(  V_{\nu,2}c_{\nu}^{\dag}c_{d_{2}}+V_{\nu,2}^{\ast}c_{d_{2}}^{\dag
}c_{\nu}\right) \nonumber\\
=%
{\textstyle\sum_{\kappa,\lambda}}
c_{\kappa}^{\dag}H_{\kappa\lambda}c_{\lambda}\nonumber
\end{gather}
Equ. (\ref{GF}) is evaluated for $\left(  \kappa,\mu\right)  $ equal
to$\ \left(  d_{1},d_{1}\right)  $, $\left(  d_{2},d_{1}\right)  $ and
$\left(  \nu,d_{1}\right)  $ where $\nu$ stands for any conduction electron state.

Together they yield%
\begin{align*}
-%
{\textstyle\sum_{\nu}}
V_{\nu,1}^{\ast}G_{\nu,d_{1}}+\left(  E+is-\varepsilon_{d_{1}}\right)
G_{d_{1},d_{1}}  &  =1\\
-%
{\textstyle\sum_{\nu}}
V_{\nu,2}^{\ast}G_{\nu,d_{1}}+\left(  E+is-\varepsilon_{d_{2}}\right)
G_{d_{2},d_{1}}  &  =0\\
\left(  E+is-\varepsilon_{\nu}\right)  G_{\nu,d_{1}}-V_{\nu,1}G_{d_{1},d_{1}%
}-V_{\nu,2}G_{d_{2},d_{1}}  &  =0
\end{align*}
From these equations one can eliminate $G_{\nu,d_{1}}$ and obtain two
equations which couple $G_{d_{1},d_{1}}$ and $G_{d_{2},d_{1}}$:%
\begin{align}
\left(  E-\varepsilon_{d,1}-%
{\textstyle\sum_{\nu}}
\frac{V_{1,\nu}V_{1,\nu}^{\ast}}{\left(  E-\varepsilon_{\nu}+is\right)
}\right)  G_{d_{1},d_{1}}-%
{\textstyle\sum_{\nu}}
\frac{V_{1,\nu}V_{2,\nu}^{\ast}}{\left(  E-\varepsilon_{\nu}+is\right)
}G_{d_{2},d_{1}}  &  =1\label{G_G}\\
-%
{\textstyle\sum_{\nu}}
\frac{V_{2,\nu}V_{1,\nu}^{\ast}}{\left(  E-\varepsilon_{\nu}\right)  }%
G_{d_{1},d_{1}}+\left[  \left(  E-\varepsilon_{d,2}\right)  -%
{\textstyle\sum_{\nu}}
\frac{V_{2,\nu}V_{2,\nu}^{\ast}}{\left(  E-\varepsilon_{\nu}\right)  }\right]
G_{d_{2},d_{1}}  &  =0\nonumber
\end{align}

For identical hopping matrix elements of impurity $d_{1}$ and $d_{2},$ i.e.
$V_{1,\nu}=V_{2,\nu}$, one can set the expression
\[%
{\textstyle\sum_{\nu}}
\frac{V_{i,\nu}V_{j,\nu}^{\ast}}{\left(  E-\varepsilon_{\nu}+is\right)
}=X=\Delta_{r}+i\Delta_{i}%
\]
equal to $X$ which consists of a real and an imaginary part.

The denominator of the two GF in (\ref{G_G}) is given by the determinant of
the matrix in (\ref{G_G}).
\[
\left(  E-\varepsilon_{d,1}-X\right)  \left(  E-\varepsilon_{d,2}-X\right)
-X^{2}%
\]
The zero points of the determinant yield the new energies of the d-levels. The
imaginary part gives the broadening of the resonance curve. These zero points
are
\[
E_{1,2}=X+\allowbreak\frac{1}{2}\left(  \varepsilon_{d,1}+\varepsilon
_{d,2}\right)  \pm\frac{1}{2}\sqrt{4X^{2}+\left(  \varepsilon_{d,1}%
-\varepsilon_{d,2}\right)  ^{2}}%
\]

The Green function $G_{d_{1},d_{1}}$ has the form%
\[
G_{d_{1},d_{1}}=\frac{1}{2}\frac{1+\frac{\left(  \varepsilon_{d,1}%
-\varepsilon_{d,2}\right)  }{E_{1}-E_{2}}}{E-E_{1}}+\frac{1}{2}\frac
{1-\frac{\left(  \varepsilon_{d,1}-\varepsilon_{d,2}\right)  }{E_{1}-E_{2}}%
}{E-E_{2}}%
\]%
\[
\]

When the two d-level energies approach each other, i.e. $\left\vert \left(
\varepsilon_{d,1}-\varepsilon_{d,2}\right)  \right\vert <2\left\vert
X\right\vert $ then one obtains%
\[
E_{1}\thickapprox2X+\allowbreak\frac{1}{2}\left(  \varepsilon_{d,1}%
+\varepsilon_{d,2}\right)  +\frac{\left(  \varepsilon_{d,1}-\varepsilon
_{d,2}\right)  ^{2}}{8X}%
\]%
\[
E_{2}\thickapprox+\allowbreak\frac{1}{2}\left(  \varepsilon_{d,1}%
+\varepsilon_{d,2}\right)  -\frac{\left(  \varepsilon_{d,1}-\varepsilon
_{d,2}\right)  ^{2}}{8X}%
\]

In this case the second term of $G_{d_{1},d_{1}}$ has a denominator with a
rather small imaginary part (proportional to $\left(  \varepsilon
_{d,1}-\varepsilon_{d,2}\right)  ^{2}$). Its resonance width is strongly reduced.

For $\varepsilon_{d,1}=\varepsilon_{d,2}$ the resonance at $E_{2}%
=\varepsilon_{d}$ has essentially no imaginary part and therefore represents a
$\delta$-function. This confirms our observation when diagonalizing the
Hamiltonian. It is the result of an interference effect between the two
d-levels by the coupling through the conduction electrons.

A sharp resonance or density of states is experimentally very desirable. It
can be used as a marker in spectroscopy. For example the sharp horn of the
superconducting BCS-density of states or the sharp Kondo resonance and others
have been used experimentally to trace energy shifts and other phenomena. The
question is whether one can fabricate a metal (or any other system) with two
resonance level which are not only identical in energy but also in their
hopping matrix elements. As one possibility I propose a one-dimensional wire
of a material with a Fermi wave length that is much larger then atomic
distances. Such a one-dimensional wire can be a narrow film of a semi metal
where the width is less then the half the Fermi wave length. This is shown in Fig.3.%

\begin{align*}
&
{\includegraphics[
height=1.0494in,
width=4.1859in
]%
{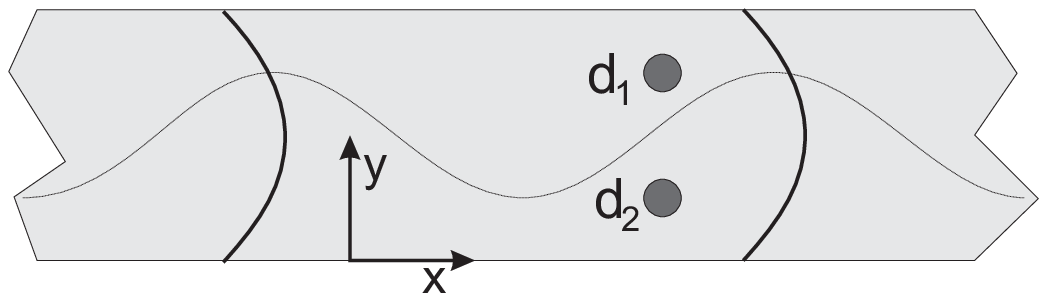}%
}%
\\
&
\begin{tabular}
[c]{l}%
Fig.3: A short part of a strip which acts as a one-dimensional\\
wire. In y-direction there is only one mode with no phase\\
modulation. The two resonance levels are inserted symmetrically\\
with respect to the middle line of the strip.
\end{tabular}
\end{align*}

Then the electronic wave functions don't change their phase in the direction
perpendicular to the wire, the phase only changes along the wire as sketched
in Fig.3. The two resonance level are inserted so that they lie symmetrically
with the central line of the wire. Then both of them are coupled with the same
hopping matrix elements to the conduction band. And both have the same d-state
energy $\varepsilon_{d}.$The distance between the two resonance levels must be
sufficiently large so that there is no direct overlap of their wave functions.

\end{document}